\documentclass[12pt]{article}
\usepackage{geometry} 
\usepackage[pdftex]{graphicx}
\usepackage{amsmath}
\usepackage{amssymb}
\usepackage{bm}
\usepackage{apacite}
\raggedright                                  

\title{Relational Dynamics in Perception:\\ Impacts on trial-to-trial variation}

\author{\normalsize{Shimon Marom}\footnote{Corresponding author (shimon.marom@gmail.com)}~~and Avner Wallach\\\textit{\normalsize{Technion - Israel Institute of Technology, Haifa 32000, Israel}}}

\date{} 

\begin{document}
\maketitle


\begin{abstract}
\noindent We show that trial-to-trial variability in sensory detection of a weak visual stimulus is dramatically diminished when rather than presenting a fixed stimulus contrast, fluctuations in a subject's judgment are matched by fluctuations in stimulus contrast. This attenuation of fluctuations does not involve a change in the subject's psychometric function. The result is consistent with the interpretation of trial-to-trial variability in this sensory detection task being a high-level meta-cognitive control process that explores for something that our brains are so used to: subject-object relational dynamics.    
\end{abstract}



\noindent Trial-to-trial variation in responses to repeated presentations of the same weak sensory object is noticeable in practically every cognitive modality.  These fluctuations, which have been labeled ``internal'', ``unexplained'' or ``inherent'' noise, are correlated over extended timescales \cite{WERTHEIMER:1953kx,Gilden:2001vn,Monto:2008ys} and inversely related to the degree of stimulus-determinability \cite{CONKLIN:1956zr}.  Over the years since the inception of psychophysics, there has been a shift in how the source of trial-to-trial variation at threshold is explained.  At present-time, the concept of noisy neural response dynamics that ``poses a fundamental problem for information processing'' is dominant \cite{Faisal:2008ly}.

Yet there is something very unnatural in the way traditional psychophysical studies of  sensory detection - studies that expose extensive trial-to-trial variation at threshold - are set up.  In real-life situations, when encountering a weak sensory stimulus that deserves attention, we try to ``do something about it''.  Consider the set of operations performed by the average man over 50 confronted with a barely detectable printed text: tilting the page, exposing it to enhanced light conditions, etc.  The stimulus itself becomes dynamic.  If the barely detected stimulus originates from another subject, we (for instance) might lean forward or ask that other person to raise his voice or to present the object in a more favorable manner.  Again, the stimulus itself becomes dynamic.  Indeed, in real-life situation, our attempts to ``do something about'' the barely detected stimulus \textit{impacts on the stimulus dynamics}, although not necessarily on our capacity to detect it.  Thus, natural perception involves an expectation of the perceiver for an ongoing coupling between his actions and the threshold-level stimulus dynamics.  In that sense, natural perception is relational.  This is real life, but in standard psychophysical experiments the situation is different:  in these experiments much effort is invested by the experimentalist to control the conditions so that the threshold-level stimulus remains static.  The subject might actively explore various features of the stimulus (``active sensing''), but any given stimulus feature in these standard psychophysical designs, remains the same regardless of the subject's behavior.  Hence no feedback between the subject's actions and the stimulus dynamics is involved, and perception becomes non-relational.

In view of the above, and encouraged by old and recent analyses that reveal rich temporal structure and non-independence in response fluctuations at threshold over extended timescales \cite[and references therein]{CONKLIN:1956zr,Gilden:2001vn,Monto:2008ys,Marom:2010ve}, we turned to examine the possibility that trial-to-trial variation in responses to repeated presentations of the same weak sensory object do not reflect an inherent noise that constrains sensory acuity and information processing.  Rather, we hypothesize that most of the observed variability in responses to weak stimuli is due to an active cognitive exploratory process, seeking for a coupling between the stimulus dynamics and subject's behavior.  To test this hypothesis we have used a generic feedback loop control algorithm, endowing a visual stimulus in a detection task the capacity to on-line match its contrast to the subject's performance, while ``clamping'' the performance at a predefined (mostly 0.5) probability of detection.  We show that once such relations are established (i.e.,~as long as the control algorithm is active), trial-to-trial variability is dramatically diminished, breaking the apparent limits of inherent noise, while keeping detection threshold and sensitivity (as reflected in the psychometric function) unchanged.  This result points at the possibility of trial-to-trial variability in sensory detection of weak stimuli being a high-level meta-cognitive control process that explores for something that life trained us to expect: subject-object relational (or, coupled) dynamics.    

\section*{Methods}

All the experiments and their analyses were performed within a Wolfram's \textit{Mathematica 7.0} environment; the software package is available on request from S.M. 

\subsection*{Psychophysical detection task}

\begin{figure}
	\centering
	\includegraphics[width=6 in]{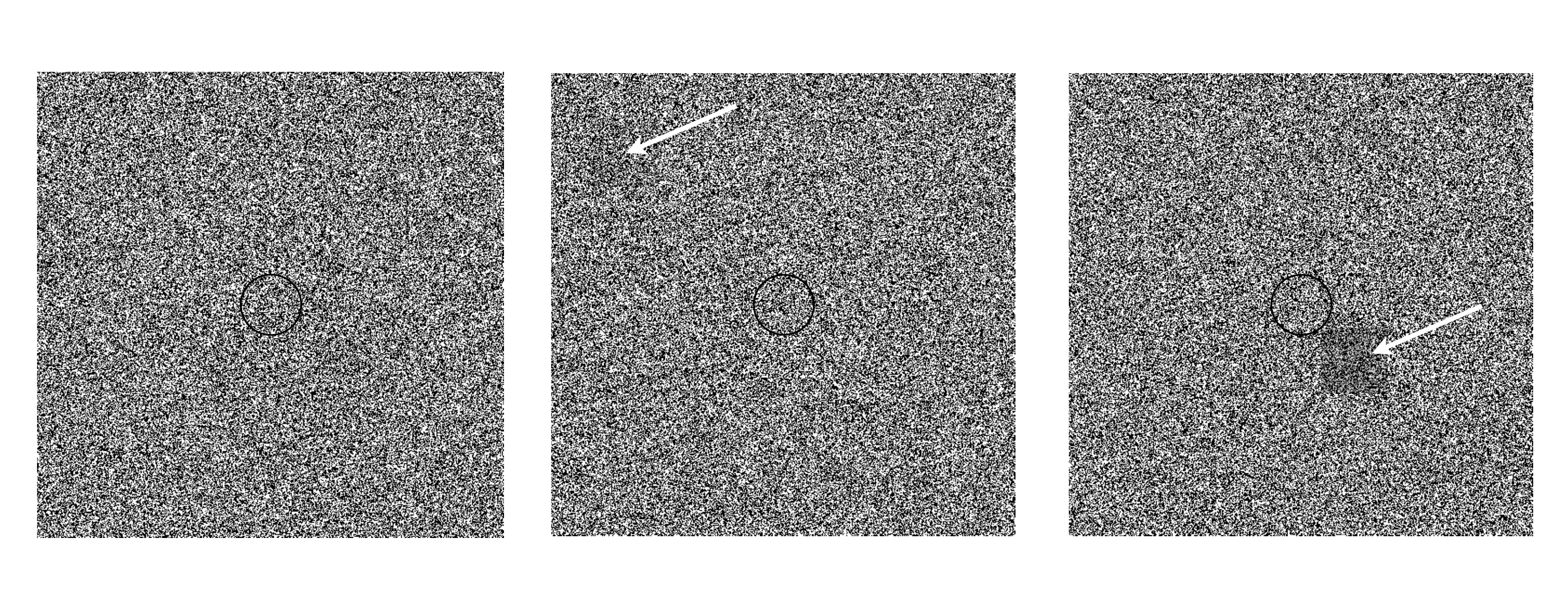} 
 	  \caption{\footnotesize{{\bf{Detection task}}.  Three examples of single trial presentations:  Background raster only (left panel), a barely detectable foreground object indicated in the top-left field of the middle panel, and an obvious foreground object (right panel). A circle at the center of the image appeared in all trials of all sessions; the subjects were instructed to try to fixate on that circle at the beginning of each trial.}}
	\label{Task}
\end{figure}

Fourteen healthy volunteers (six females), graduate students and post-docs at the age of 27-40 year, were the subjects of this study.  Unless indicated otherwise, the basic visual detection task is as follows (see Figure \ref{Task}):  A random 500x500 background raster of black and white pixels, occupying 135x135 millimeters, was presented in the center of a flat Apple 24 inch screen.  A single session was composed of 500 presentation trials of the raster, randomized in each trial.  The raster remained on screen for half a second in each trial.  A smaller foreground raster of 70x70 randomized grayscale pixels was embedded within the background raster area.  The gray-level (denoted $x$) of the $i$-th pixel in the $n$-th trial was determined by a uniformly distributed random number ($0\leq r_{i,n} \leq 1$) such that  
$$
x_{i,n} = \left\{ 
  \begin{array}{l l}
    0 \text{ (white)} & \quad \text{for $r_{i,n} < 0.5$}\\
    1 \text{ (black)} & \quad \text{for $r_{i,n} > C_{n}$}\\
    r_{i,n}  & \quad \text{for $0.5 \le r_{i,n} \le C_{n}$}\\
  \end{array} \right.
  $$
where $C_{n}$, referred to as ``contrast'', was calculated on a trial-by-trial basis as described in the next section.  With this procedure the general pattern of background black-and-white scatter is present also within the foreground, while the range $[0.5,C_{n}]$ of foreground grayscale serves as a control variable.  Of the 500 trials in a session, 50 randomly introduced sham trials did not include any foreground object.  The position of the foreground object in each trial was randomized.  After a trial, subjects were asked to press one of two keys, signifying whether they detected or not the foreground object.  Each trial immediately followed the subject's response to the preceding trial; no time limit was set for the subject to produce an answer. 

\subsection*{Stimulus control algorithm}

The experimental design was adopted from a recently introduced Response Clamp methodology for analysis of neural fluctuations (Wallach et al., \textit{Arxiv preprint arXiv:1008.1410}), with modifications that enabled its application to the present behavioral setting.  A Proportional-Integral-Derivative (PID) controller was realized in Wolfram's \textit{Mathematica 7.0} environment.  The input to the controller is the error signal,
\[e_{n}=P_{n}^{*}-\widetilde{P}_{n}\]  where $P_{n}^{*}$ and $\widetilde{P}_{n}$ are the desired and actual detection probabilities (calculated as explained below) at the $n$-th trial, respectively. The output of the controller is generally composed of three expressions,

\[y_{n}=g_{P}e_{n}+g_{I}\sum_{i=0}^{n}e_{i}+g_{D}(e_{n}-e_{n-1})\]
 where $g_{P},g_{I}$ and $g_{D}$ are the proportional, integral and derivative gains, respectively; $g_{P}$ was set to 1.0, $g_{I}$ to 0.02, and $g_{D}$ to either 0.02 or 0 (with no appreciable effect). Finally, the contrast $C_{n}$ equals the controller's output plus some baseline: \[C_{n}=y_{n}+C_{baseline},\] where $C_{baseline}=0.5$. 

\subsection*{Calculation of detection probability}

Response probability was estimated on-line as follows:  Let $s_{n}$ be an indicator function, so that $s_{n}=1$ if the subject detected the $n$-th foreground stimulus and $s_{n}=0$ otherwise. We define $\pi(n)$ as the probability of the subject to detect a foreground stimulus at trial $n$. We can estimate this probability using all past responses  $\{s_{i}\}_{i=1}^{n}$, by integrating them with an exponential kernel, 
$$
\widetilde{P}_{n}=\widetilde{P}_{0} e^{-\frac{n}{\tau}}+\sum_{i=1}^{n}s_{i}(1-e^{-\frac{1}{\tau}})e^{-\frac{n-{i}}{\tau}}\text{,}
$$
where $\tau$ is the kernel's decay-constant. To compute this on-line, we used the recursive formula:
$$
\widetilde{P}_{n}=s_{n}(1-e^{-\frac{1}{\tau}})+\widetilde{P}_{n-1}e^{-\frac{1}{\tau}}\text{,}
$$
setting $\tau=10$ trials, and $\widetilde{P}_{0}=0.5$.

\subsection*{Closed-Loop, Replay and Fixed contrast modes}

In the basic design, each subject was exposed to three experimental sessions denoted \textit{closed-loop}, \textit{replay} and \textit{fixed}.  The first session was always a \textit{closed-loop} session, whereas the second and third were \textit{replay} and \textit{fixed} sessions, introduced in an alternating order to different subjects.  A 10 minutes break was given after the first and second sessions.  

In the \textit{closed-loop} session, the desired response probability ($P_{n}^{*}$) was kept constant ($P_{n}^{*}= 0.5$, unless indicated otherwise in the main text) and the control algorithm operated as explained above, updating the contrast ($C_{n}$) of the foreground object from one trial to the next based on the error signal ($e_{n}$). The series of 450 $C_{n}$ values produced in this \textit{closed-loop} session (500 trials minus the 50 sham trials), served for the generation of the foreground objects in the \textit{replay} session.  Thus, in the \textit{replay} session the control algorithm was disconnected, yet we were able to record the responses of the subject to exactly the same series of contrasts presented in the \textit{closed-loop} session, but now in an open-loop context; detached from the trial-by-trial coupled observer's - observed dynamics.  In the \textit{fixed} session, the average contrast calculated from the series of above mentioned contrasts was used for all presentations, thus omitting stimulus variance altogether.  This \textit{fixed} session allowed us to estimate the impact of stimulus fluctuations on response dynamics.

\section*{Results}

The nature of the detection task is demonstrated in Figure \ref{Task}:  The left panel shows a background raster only.  The middle panel shows a barely detectable foreground object (indicated in the top-left field).  The right-hand panel demonstrates an obvious foreground object. The probability of false positive detection, calculated from responses of all subjects to the 50 sham trials was negligible (.017, SD $=.029$, $n=8$), indicating that the subjects did not tend to report detection when they did not really see something.  The average response time was around 1 second per trial, slightly longer in the \textit{closed-loop} session (1.03 sec, SD $=1.3$) compared to \textit{replay} and \textit{fixed} sessions (0.89 sec, SD $=1.6$ and 0.81 sec, SD $=1.0$, respectively).  Response time distributions in all three sessions had a long right tail (coefficient of skewness: 11, 17 and 19, respectively).

\begin{figure}
	\centering
	\includegraphics[width=5 in]{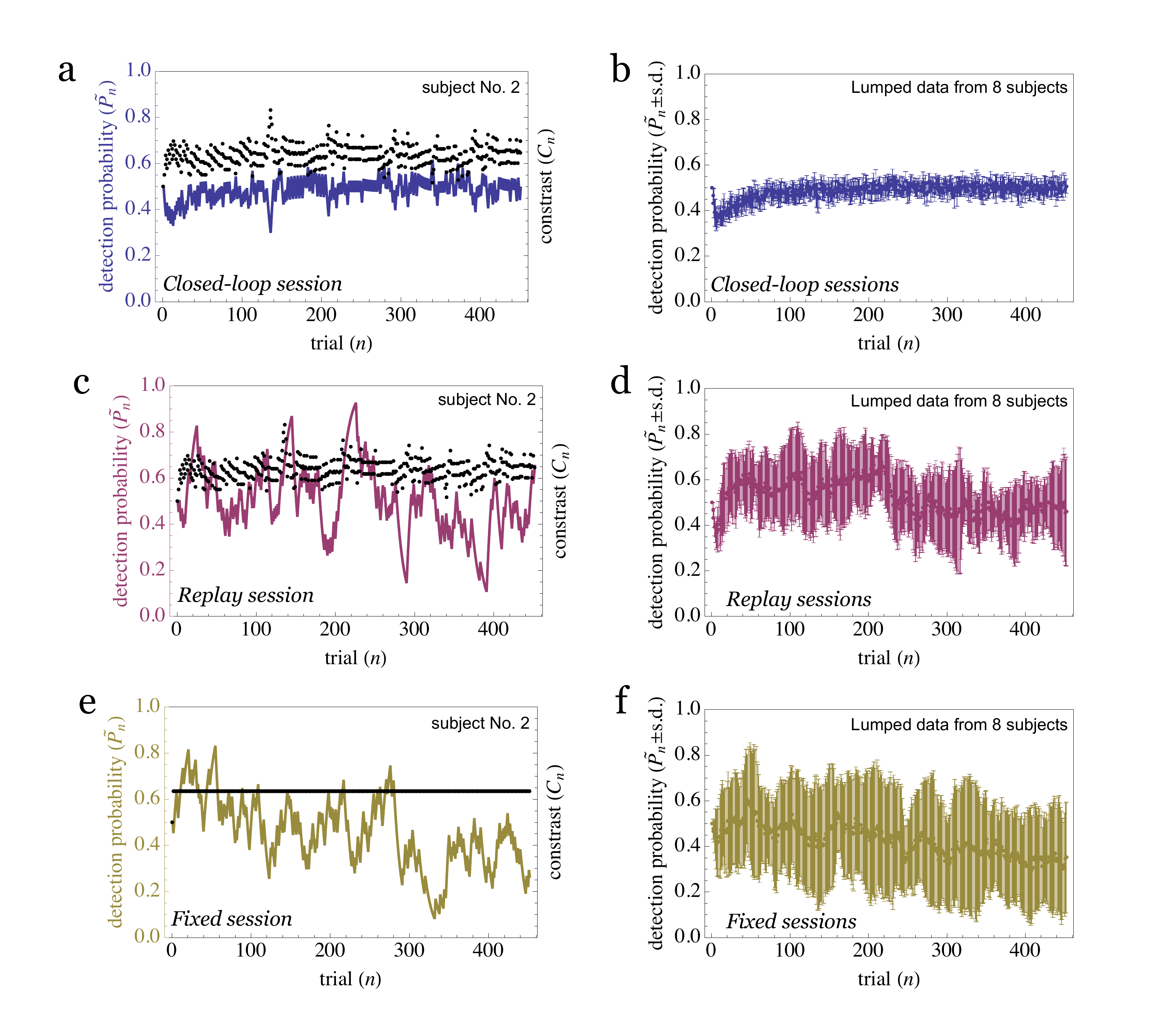} 
 	  \caption{\footnotesize{ {\bf{Detection probability in \textit{closed-loop}, \textit{replay} and \textit{fixed} sessions.}}  Data obtained from an experiment on one individual (left column) and the summation of observations from the eight different subjects that were tested in this protocol (right column).  In all cases, the fluctuations around mean detection probability are significantly smaller in the \textit{closed-loop} session.  The initial decline in detection probability, which is most apparent in panels (a-d) reflect the initial setting of both $\widetilde P_{n}$ and $C_{n}$ to 0.5.  Error bars in the righthand column depict standard deviation across all subjects.}}
	\label{MainObservation}
\end{figure}

The main observation is shown in Figure \ref{MainObservation}, where the responses of one subject (left column) and the group of eight subjects (right column) that were tested in \textit{closed-loop} (dark blue, top row), \textit{replay} (purple, middle row) and \textit{fixed} (a kind of yellow, bottom row) sessions are plotted.  Let us start with panel (a) of Figure \ref{MainObservation}, describing the results obtained in the \textit{closed-loop} session of one individual.  The desired detection probability $P_{n}^{*}$ (see methods) was set to 0.5, and the control algorithm updated the contrast $C_{n}$, trial-by-trial, as indicated by black dots (righthand y-axis of Figure \ref{MainObservation}(a)). The resulting estimated detection probability ($\widetilde{P}_{n}$, dark blue, lefthand y-axis) of that individual gradually approached the desired value,  albeit fluctuating about it.  Also note the expected anti-correlation of $C_{n}$ and $\widetilde{P}_{n}$.   Panel (b) of Figure \ref{MainObservation} shows the average performance  (and standard deviation) of all eight subjects that participated in such a \textit{closed-loop} session, showing that the controller converges within ca.~100 trials, and succeeds in ``clamping'' the detection probability at around 0.5, as preset.

Figure \ref{MainObservation}(c) shows the performance of the same individual whose data is shown in panel (a), but now in the \textit{replay} session, where the $C_{n}$ series obtained in the \textit{closed-loop} session (black dots) is ``replayed'', regardless of the subject's responses.  Under these conditions the control algorithm is shut down and stimulus contrast is completely decoupled from the subject's behavior.  Note the emergence of large slow fluctuations around the preset 0.5 detection probability; despite the fact that the stimulus series is practically identical to that shown in panel (a), the performance in (c) is very different.  Panel (d) of Figure \ref{MainObservation} shows the average performance  (and standard deviation) of all eight subjects that participated in the \textit{replay} session.    

And finally, as demonstrated in Figure \ref{MainObservation}(e), when the average $C_{n}$ of the subject - whose data is shown in panels (a) and (c) - is used in the \textit{fixed} session, where all 450 trials have identical contrast (black line of panel (e)), large slow fluctuations and drift are observed.   Panel (f) of Figure \ref{MainObservation} shows the average performance (and standard deviation) of all eight subjects that participated in these \textit{fixed} session.   

\begin{figure}
	\centering
	\includegraphics[width=5 in]{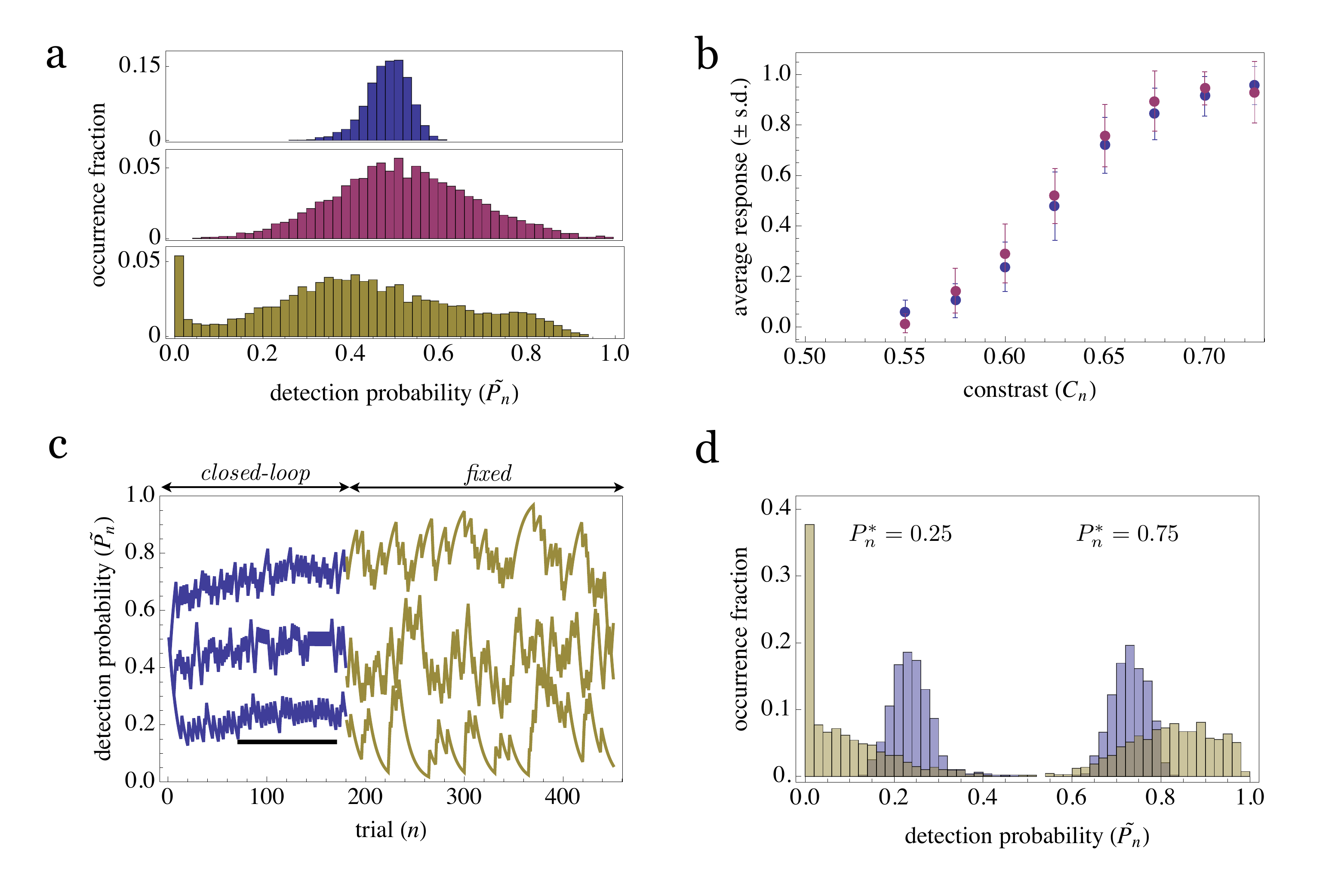} 
 	  \caption{\footnotesize{{\bf Psychometrics and dependence of fluctuations on subject's performance.}  (a) Histograms of detection probability ($\widetilde P_{n}$), calculated from data of all subjects, in \textit{closed-loop} (top), \textit{replay} (middle) and \textit{fixed} session. (b) Both threshold and sensitivity are practically identical in \textit{closed-loop} (dark blue) and \textit{replay} (purple) sessions.  The curves were calculated by averaging the responses (1's and 0's), for each of the eight subjects, in different contrast ($C_{n}$) bins; binsize $=.025$.  A minimum of 5 occurrences of a given contrast per bin was set as a requirement for inclusion in the calculation.  The \textit{Average Response} (y-axis) and its standard deviation among subjects for each contrast bin are shown in the plot.  Note that average response thus calculated is not the same quantity as detection probability $\widetilde P_{n}$ (the latter takes into account the \textit{temporal order} of responses).   (c) A \textit{closed-loop} mode is instantly switched to a \textit{fixed} mode, by disconnecting the controller and using a constant contrast value (average of the  $C_{n}$ series over a time segment depicted by a black bar).  Data obtained from a single subject in three different preset  $P^{*}_{n}$  values (0.25, 0.5 and 0.75).  (d)  Average distributions of detection probability ($n=4$ subjects) for $P^{*}_{n}=0.25$ and $P^{*}_{n}=0.75$ sessions as shown in (c);  obtained separately from the \textit{closed-loop} (blue) and \textit{fixed} phases.}}
	\label{Overall}
\end{figure}

The group statistics of Figure \ref{MainObservation}(b, d and f) are summarized in Figure \ref{Overall}(a).  Clearly, the best performance is obtained when relational dynamics are allowed between $\widetilde P_{n}$ (the performance of the observer) and $C_{n}$ (the contrast series).  One possible explanation to this result is that under these different experimental conditions there is a change in the sensitivity of the subject to the stimulus.  Figure \ref{Overall}(b) shows the psychometric functions for \textit{closed-loop} and \textit{replay} sessions, calculated by averaging the responses (1's and 0's), for each of the eight subjects, in different contrast ($C_{n}$) bins.  (Note that average response thus calculated is not the same quantity as detection probability $\widetilde P_{n}$; the latter takes into account the \textit{temporal order} of responses.)  Clearly, these psychometric functions show that both threshold and sensitivity extracted from the responses of all the subjects, are practically identical in \textit{closed-loop} and \textit{replay}.  However, the richness of the dynamics and the marked differences between \textit{closed-loop} and \textit{replay} modes seen in Figure \ref{MainObservation}, are practically averaged out when the data are collapsed to standard psychometric functions of the kind shown in Figure \ref{Overall}(b). 

The importance of instantaneous coupling between the observer's behavior and the stimulus dynamics is demonstrated in Figure \ref{Overall}(c), where a \textit{closed-loop} mode is instantly switched to a \textit{fixed} mode, by disconnecting the controller and using a constant contrast value (average of the  $C_{n}$ series over a time segment depicted by a black bar).  Figure \ref{Overall}(c) shows data of a single subject, in three different preset  $P^{*}_{n}$  values (0.25, 0.5 and 0.75).  As soon as the coupling of observer's-observed dynamics is disconnected, the variance of detection probability markedly increases and slow correlations seem to emerge, at all $P^{*}_{n}$ tested.  Interestingly, as shown in the averaged histograms of Figure \ref{Overall}(d), within the \textit{fixed} phase of the experiment of panel (c) the detection probabilities seem to have a ``binary'' preference towards 1 or 0 (for the cases of $P^{*}_{n}=0.75$ and $P^{*}_{n}=0.25$, respectively).  This preference stands in contrast to the symmetric case of $P^{*}_{n}=0.5$ (e.g. Figure \ref{Overall}(a)).

\section*{Concluding remarks}

Two basic observations are presented here.  The \textit{first} is that trial-to-trial variability in sensory detection of a weak visual stimulus is dramatically diminished when rather than presenting a stimulus contrast that is independent of the subject's ongoing actions, fluctuations in a subject's judgment are matched by fluctuations in stimulus contrast.  Clearly, this result reaffirms that trial-to-trial fluctuations are not ``noise'' in the strict sense of being independent of each other.  Moreover, the significant difference, between features of fluctuations measured when dynamic observer-observed relations exist, and those measured in the absence of such coupled dynamics, calls for re-examination of the way psychophysical experiments are conducted.   Indeed, measuring temporal fluctuations of a psychophysical function under open-loop conditions, where there is no relation between subject performances and sensory object contrast dynamics, is a most un-natural setting.  Here we implemented an adaptive algorithm (PID) borrowed from control theory in order to couple the observer-observed dynamics.  While the PID control algorithm has theoretical advantages in the present context, there exist many other adaptive psychophysical procedures \cite{Treutwein:1995qf} that are, actually, in use when experimentalists attempt to identify points of interest on psychometric functions.  We propose to substantially extend their use in order to expose the dynamics of perception under more natural  experimental conditions.  The \textit{second} basic observation is that the above diminishing of trial-to-trial fluctuations by coupling between observer-observed dynamics, is not accompanied by a change in sensory sensitivity to the input.  Taken together, the two basic observations suggest that trial-to-trial variability in sensory detection of weak stimuli might reflect a high-level control process. 

As pointed out by Wertheimer (1953), trial-to-trial variation at threshold was generally attributed, in the early days of psychophysics, to uncontrolled experimental conditions, with the assumption that the subject is stable.  Response fluctuations, however, were soon shown to be non-independent \cite{verplanck1952nonindependence}; that is - successive responses to repeated presentations of the same threshold stimulus (in auditory, visual and somatosensory modalities) are correlated over timescales ranging from seconds to days \cite{WERTHEIMER:1953kx}.  These long range trial-to-trial response correlations were then interpreted as reflecting a meta-cognitive \textit{guessing} process that is active where there is no possibility for stimulus-determination; the assumption being that ``guesses are more likely to be influence by preceding responses (success and failures) than are sensory judgments'' \cite{CONKLIN:1956zr}.  Later years brought with them a more reductionistic focus on neural sources of ``noise'' and ``short-term plasticity'' that may account for observed trial-to-trial response variability \cite{Faisal:2008ly}.  Viewed from this historical angle, the results presented here pull the pendulum back to the meta-cognitive pole, offering an interpretation according to which - when facing a weak stimulus - subjects vary their response patterns, seeking to establish  \cite<predictive? see>{Rosen:1985fk,Creutzig:2009fk} relations between their actions and  the dynamics of stimulus features.  This interpretation, in its broader sense, goes far beyond  psychophysics of weak stimulus detection; it touches upon what psychologists try to tell us over the past fifty years on the developing mind \cite<e.g.>{Stern:1985fk}, when we care to listen.

\bibliographystyle{apacite}  
\bibliography{PsyRC}
\end{document}